\PassOptionsToPackage{table}{xcolor}
\documentclass[trackchanges, twocolumn]{aastex701}

\usepackage{graphicx}
\definecolor{lightBlue}{HTML}{6baed6}  
\definecolor{darkBlue}{HTML}{2171b5}   
\definecolor{midBlack}{HTML}{000000}   
\definecolor{darkRed}{HTML}{cb181d}    
\definecolor{lightRed}{HTML}{ff6666}   
\usepackage{hyperref} 

\begin{document}


\title{Very Massive, Rapidly Spinning Binary Black Hole Progenitors \\ through Chemically Homogeneous Evolution -- The Case of GW231123 \\
}

\author[orcid=0009-0006-3186-5826, gname=Silvia A., sname=Popa]{Silvia A. Popa}
\affiliation{Max Planck Institute for Astrophysics, Karl-Schwarzschild-Str.~1, 85748 Garching, Germany}
\email[show]{spopa@mpa-garching.mpg.de}

\author[orcid=0000-0001-9336-2825]{Selma E. de Mink}
\affiliation{Max Planck Institute for Astrophysics, Karl-Schwarzschild-Str.~1, 85748 Garching, Germany}
\email{}


\begin{abstract}

Among the over 200 gravitational wave detections reported so far, GW231123 is a remarkable event that not only hold the record for the most massive black hole merger, but also exhibits extreme spins. Its origin is actively debated. Proposed scenarios include dynamical formation involving a sequence of mergers, Population III stars, accretion in an AGN disk and also more exotic explanations including primordial black holes and cosmic strings, each facing different challenges. Recent work showed that the incoming black holes of GW231123 can be formed out of massive rapidly rotating collapsing helium stars.  
Here, we address the question  how such very massive rapidly rotating helium stars can be formed in very close binary systems. For this we explore chemically homogeneous evolution (CHE) involving progenitors with masses above the pair-instability mass gap. We compute a grid of detailed massive binary models with the stellar evolution code MESA to follow the early evolution of binary progenitors and show that: (i) very massive ($M_i > 140\, M_\odot$) CHE binaries at low metallicity ($Z=10^{-5}$) naturally produce rapidly rotating progenitors with high masses and high spins matching the properties of the black holes in GW231123 and (ii) the maximum spin of the progenitors is bound by their critical rotation rate leading to a tight correlation between the dimensionless spin and mass, $a \propto M^{-0.9}$, in models that have no hydrogen left. We conclude that the CHE channel appears to be a viable and natural scenario to produce progenitors. We compare and discuss the differences with earlier studies and comment on the large uncertainties in the final fate and collapse.

\end{abstract}
\section{Introduction}

Recently, the LIGO--Virgo--KAGRA collaboration reported the detection of GW231123, the most massive binary black hole merger observed to date \citep{GW231123}. The event involved two black holes with masses of \(137^{+22}_{-17}\,M_\odot\) and \(103^{+20}_{-52}\,M_\odot\), which places both components within or above the predicted pair-instability mass gap \citep{Heger_2002, Farmer_2019, Farmer_2020, Woosley_2021, Farag_2022, Mehta_2022}. Both black holes also exhibit exceptionally high spins of \(\chi_1 = 0.9^{+0.10}_{-0.19}\) and \(\chi_2 = 0.8^{+0.20}_{-0.51}\), the largest spins observed so far in a gravitational-wave (GW) event. 

Since its announcement, GW231123 has inspired several theoretical studies attempting to explain its origin and formation history. Proposed explanations range from astrophysical formation channels to more exotic scenarios, such as primordial black holes formed in the early universe \citep{Yuan_2025, De_Luca_2025} and cusps or kinks on a cosmic string \citep{Cuceu_2025}.

Several authors considered hierarchical assembly through dynamical formation, where black holes (BHs) grow in mass via successive mergers in dense stellar environments \citep[e.g.,][]{Miller_2002, Fischbach_2017, Yang_2019, Gerosa_2021}. \cite{Li_2025} suggest that the progenitors of GW231123 originated from the hierarchical merger of 4 and 6 first-generation black holes to make the primary and secondary component. To prevent the intermediate merger products from escaping due to the recoil kicks, this channel requires an environment with sufficiently deep potential wells, such as those found in nuclear star clusters or Active Galactic Nuclei (AGN). \cite{Delfavero_2025} explore the AGN formation channel and propose that GW231123 is most consistent with a merger of third- or fourth-generation BHs.  

 \cite{Stegmann_2025}, however, point out that the very high spins in the progenitors are very unlikely in dynamically assembled systems. Due to the isotropic distribution of the spin orientations of the incoming black holes, a spin distribution peaking at 0.7 is expected \citep{Gerosa_2017, Fischbach_2017}.  They argue that spins above $\chi_{i} > 0.9$ can only be formed through mergers of first-generation black holes that already have large and closely aligned spins pointing toward an isolated binary evolution channel, such as the chemically homogeneous evolution channel, as a necessary precursor.

Alternatively, \cite{Bartos_2025} propose that the extreme properties of GW231123 might be explained by sustained accretion, which may occur either in AGN disks or in metal-poor or metal-free Population III stars.  Indeed, spin up may be expected as a result of accretion in the case of a thin disk \citep{Bardeen_1970, King_Kolb_1999}. However, at super-Eddington accretion rates, the inflow is expected to form a thick accretion disk or even reach a near-radial geometry \citep{Volonteri_2005}. It is debated whether the black hole would spin up under such conditions. \cite{Tchekhovskoy_2012} even show how, under the influence of magnetic fields, the black hole may even spin down.

\cite{Tanikawa_2025} instead propose an isolated binary evolution channel involving Population III stars. This channel relies on several assumptions. The progenitor stars must have very small cores to avoid disruption induced by pair-instability, and they should have very massive envelopes, which are retained during stellar evolution and subsequently accreted onto the resulting black holes.

Finally, \citet{Croon_2025} and \citet{Gottlieb_2025} explore the evolution and collapse of very massive, rapidly rotating single helium cores above the mass gap as potential progenitors of the incoming black holes of GW231123. \citet{Croon_2025} follow the evolution using the 1D stellar evolution code MESA until core collapse and estimate the final black hole mass following \citet{Batta_2019}. \citet{Gottlieb_2025} start from a MESA model of a helium core and use general-relativistic magnetohydrodynamic simulations to model the collapse, BH formation and the subsequent accretion. Both studies find solutions consistent with the inferred masses and spins of GW231123’s incoming black holes.

In this study, we explore how such very massive rapidly rotating helium stars may form in a binary system through the chemically homogeneous evolution (CHE) channel. This channel has been proposed to produce nearly equal-mass, high-spin black hole binaries that merge within a Hubble time in low-metallicity environments \citep{Mink_2016, Mandel_2016, Marchant_2016, Buisson_2020}. Here, we focus on progenitors with masses above the pair-instability mass gap, capable of producing black holes consistent with the properties inferred for GW231123.

The CHE channel consists of massive, rapidly rotating stars in tight, initially tidally-locked binaries that can avoid the post-main-sequence expansion expected from classical stellar evolution due to efficient chemical mixing \citep{Paczynski_1976, Maeder_1987, Yoon_2006}. Rotation and tidal deformation induce steady-state currents in these stars that transport helium to the surface and fresh hydrogen to the core, leading to an almost chemically homogeneous atmosphere \citep{DeMink_2008, DeMink_2009, Hastings_2020}. Even though the efficiency of rotational mixing is debated, CHE is expected to occur naturally in very massive stars during central hydrogen burning, as their convective cores encompass $80 - 90 \%$ of their mass.

Chemically homogeneous massive stars remain compact, evolve into helium stars, and may later collapse into black holes. In low metallicity environments, mass loss due to stellar winds is minimal, allowing such binaries to retain their tight separation throughout their evolution. Consequently, the final black holes are close enough for GW emission to drive a merger within the lifetime of the Universe. Low mass loss also means the stars preserve a large fraction of their initial angular momentum, which in the case of complete collapse, leads to black holes with high spins - making CHE a promising channel to reproduce the high spins inferred for GW23123's progenitors. 

This work is organized as follows. In Section \ref{sec:methods}, we discuss our method and review the physical assumptions. Section \ref{sec:results} presents our grid of CHE systems. We discuss our results and their implications in Section \ref{sec:discussion}, followed by the conclusions in Section \ref{sec:conclusions}.

\section{Methods}
\label{sec:methods}

We use the open-source one-dimensional stellar evolution code MESA, version r23.05.1, \citep{Paxton_2011, Paxton_2013, Paxton_2015, Paxton_2018, Paxton_2019, Jermyn_2023}. Our work largely builds on the methodology and physical implementation presented in \citet{Marchant_2016}, see also \cite{Buisson_2020} and \cite{Sharpe_2024}. Below, we summarize the main assumptions. 


\hspace{0.5cm}\textit{Internal Mixing and Angular Momentum Transport.}
The very massive stars considered in this work develop extended convective interiors during central hydrogen burning, which is the most important process in keeping 80-90\% of their interior chemically homogeneous. Convection is modeled using the standard mixing-length theory, adopting the Ledoux criterion, with a mixing-length parameter $\alpha = 1.5$ \citep{Bohm-Vitense_1958}. Convective core overshooting is included during core hydrogen burning with a step function of $\alpha_\mathrm{OV} = 0.345$ \citep{Brott_2011}. We also consider semiconvection, with an efficiency parameter $\alpha_{sc} = 1.0$ \citep{Langer_1983}.

We further consider rotationally-induced mixing to operate in layers that are not yet mixed by convection. The main process in our models is Eddington-Sweet circulation, but we also include contributions from secular and dynamical shear instabilities, and the Goldreich-Schubert-Fricke (GSF) instability. The overall mixing efficiency is set by the parameter $f_c = 1/30$ \citep{Pinsonneault_1989, Chaboyer_1992, Heger_Langer_Woosley_2000, Brott_2011}.  To account for enhanced mixing due to tidal deformation in close binaries, we enhance the diffusion coefficient of the Eddington-Sweet circulation to $D_\mathrm{ES} = 2.1$, following \cite{Hastings_2020} as adopted by \cite{Sharpe_2024}.

We further account for angular momentum transport as a result of magnetic fields.  For this, we adopt the widely-used Spruit-Tayler dynamo \citep{Spruit_2002}. Even though the validity of the precise implementation is debated \citep{Denissenkov_2007, Zahn_2007}, the need for internal AM transport is supported by observations \citep{Suijs_2008, Fuller_2015}.  In practice, our models stay near rigid rotation until the end of helium burning.

\textit{Tidal Interactions.}
During the main sequence evolution, where the radii of the stars are large compared to the orbit, we account for tidal interactions following the formalism of \cite{Zahn_1975, Zahn_1977} for stars with radiative envelopes. The tidal synchronization timescale $\tau_{\mathrm{synch}}$ is given by: 
\begin{equation}
\small
\frac{1}{\tau_{\mathrm{synch}}}
\propto  E_2 
\left( \frac{GM}{R^{3}} \right)^{1/2} 
\frac{M R^{2}}{I} 
q^{2} (1 + q)^{5/6} 
\left( \frac{R}{a} \right)^{17/2} ,
\label{eq:tides}
\end{equation}
where $M$ is the stellar mass, $R$ the stellar radius, $I$ the stellar moment of inertia, $q = M_2/M_1$ the mass ratio, $a$ the orbital separation, $G$ the gravitational constant, and $E_{2}$ the second-order tidal coefficient, which depends on the stellar structure. For the latter, we use a simple fit by \citet{Hurley_2002}. Although this fit is strictly only valid for zero-age main-sequence stars, we find that this approach effectively keeps the stars synchronized during central H burning as expected for near contact systems.   

For post-main sequence evolution, the expression for E2 is no longer valid \citep[e.g.\ ][]{Qin_2022} and detailed simulations suggest that the expression above may overestimate the efficiency of tides in this phase \citep{Ma_Fuller_2023}. In our simulations, we account for this by assuming that tides become inefficient after the end of hydrogen burning. We deem this reasonable, since at this stage, the stars contract into compact helium stars, with their radii decreasing by a factor of 2.5. This would lower the efficiency of tides by 3 orders of magnitude, given the very strong dependence of the synchronization timescale on the radius ($t_\mathrm{sync} \propto R^{9}$, see Eq.~\ref{eq:tides}). In practice, we implement this once the central hydrogen abundance drops below $X_\mathrm{c} < 0.005$.

\textit{Stellar Winds and Mass Loss.}
Stellar winds are implemented following the prescription of \cite{Yoon_2006}, as adopted by \cite{Brott_2011}. For hydrogen-rich stars with surface helium abundance $Y_s < 0.4$, the wind recipe is taken from \cite{Vink_2001} while for hydrogen-poor stars ($Y_s >0.7$) the scheme from \cite{Hamann_1995} is used, reduced by a factor of 10. For stars with intermediate helium surface abundances ($0.4 < Y_s < 0.7$), the mass-loss rate is interpolated between the two regimes. 

In all cases, a metallicity scaling of $(Z/Z_\odot)^{0.85}$ is applied \citep{Mokiem_2007}. In our models, we adopt a very low metallicity of $Z = 10^{-5} = 0.003 \cdot Z_\odot$, using a solar metallicity of $Z_\odot = 0.0142$ as in \cite{Asplund_2009}, to minimize the mass and angular momentum loss via winds.

Since our models involve rapidly rotating stars, we also include rotationally enhanced mass-loss as in \cite{Heger_2000} (see also \cite{Aguilera-Dena_2018}). For stars approaching critical rotation ($\Omega/\Omega_\mathrm{crit} > 0.98$), MESA implicitly adjusts the mass-loss rate to ensure that the rotation rate remains just below the threshold. The critical rotation rate, $\Omega_\mathrm{crit}$, is determined by the balance between the centrifugal and the gravitational force at the stellar equator, taking into account the reduction in effective gravity due to radiation pressure, quantified by the Eddington factor $\Gamma$, and is given by:
\begin{equation}
\small
\Omega_\mathrm{crit} = \sqrt{\frac{GM(1-\Gamma)}{R_\mathrm{eq}^3}} \mathrm{ \,\,\, with} 
\quad \Gamma = \frac{L}{L_\mathrm{edd}} = \frac{\kappa L}{4\pi c G M},
\label{eq:omega_crit}
\end{equation}
where $M$ is the stellar mass, $R_\mathrm{eq}$ the equatorial radius, $L_\mathrm{edd}$ the Eddington luminosity, $\kappa$ the opacity, and $c$ the speed of light.

\textit{Overcontact Systems.}
Some binary systems are in such a tight orbit at the onset of hydrogen burning that both stars are overflowing their Roche volumes. Here, we follow \cite{Marchant_2016}
to model the treatment of such over-contact systems. If the stars overflow the outer Lagrangian point L2, mass loss and large angular momentum loss are expected to drive the two stars to merge.  We stop the evolution of such systems.

\textit{Reaction Rates and Microphysics.}
We use a simple nuclear network \texttt{basic.net}, which contains 8 isotopes, which is sufficient for the burning stages considered in this work. Nuclear reaction rates are taken from \citet{Angulo_1999} and \citet{Cyburt_2010}, with screening accounted for as in \citet{Chugunov_2007} and neutrino emission as determined in \citet{Itoh_1996}. Opacities were computed from the OPAL project \citep{Iglesias_1996}. The equations of state for different temperatures, densities and compositions are determined from a combination of \citet{Saumon_1995}, \citet{Timmes_2000}, \citet{Rogers_2002}, \citet{Potekhin_2010}, \citet{Irwin_2012} and \citet{Jermyn_2021}.

\textit{Assumptions for the Final Mass and Spin of the Black Holes.}
Since the focus of this work is the early evolution of the progenitor systems, we only evolve our models until central helium depletion. We estimate its further fate based on earlier MESA simulations by \citet{Farmer_2019, Farmer_2020} and \citet{Mehta_2022} and assume helium cores more massive than $120\, M_\odot$ to form black holes above the pair-instability mass gap. This corresponds to the outcome of models assuming a slightly enhanced $^{12}\mathrm{C}(\alpha,\gamma)^{16}\mathrm{O}$ reaction rate $1$–$3\sigma$ stronger than the canonical value.

\cite{Marchant_2020} explore the impact of rotation on the pair-instability mass gap and find that the effect is very small, changing the location of the lower edge by only 4\% in simulations that account for efficient internal angular momentum transport. We assume this also holds for the upper edge of the mass gap and consider this effect small enough to ignore, see however \citet{Croon_2025}.  

In this work, we quote the final masses and dimensionless spins at the end of our simulation. The dimensionless spin, or Kerr parameter, $a_\mathrm{}$ is computed as:
\begin{equation}
    a_\mathrm{} = \min\left(1, \frac{c\,J}{G\,M^2}\right),
    \label{eq:a_BH}
\end{equation}
where $J$ and $M$ are the star's angular momentum.  Determining the actual final mass and spin depends on the collapse and the detailed physics of accretion and is subject to very substantial uncertainties, see Sect.~\ref{sec:discussion}.  However, our approach allows us to place an upper limit to the mass and angular momentum of the black hole that may be formed.  In the case where further mass and angular momentum loss is negligible, our final spins and masses would correspond to the individual black hole masses and spins or Kerr parameters, assuming a direct collapse and neglecting energy losses from neutrino emission as they are expected to be small \citep{Zevin_2020}. 

\begin{figure}
    \centering
    \makebox[\linewidth]{%
    \includegraphics[width=1.07\linewidth]{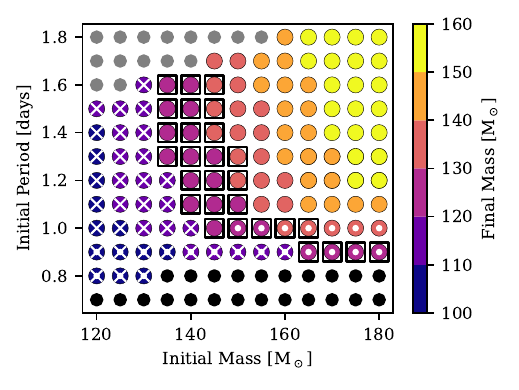}
    }
    \caption{Grid of binary systems showing the different evolutionary outcomes: chemically homogeneously evolving (CHE) systems (color-coded by the resulting black hole (BH) mass assuming direct collapse), non-CHE systems (gray) and systems that overflow their L2 point on the main sequence (black). We mark overcontact systems at the onset of hydrogen burning by white dots and systems that leave no remnant behind due to pair instability by white crosses. Highlighted by the black squares are systems where our final models have masses and spins consistent with GW231123’s progenitors.}
    \label{fig:grid_final_mass}
\end{figure}

\begin{figure*}
    \centering
    \includegraphics[width=0.85\linewidth]{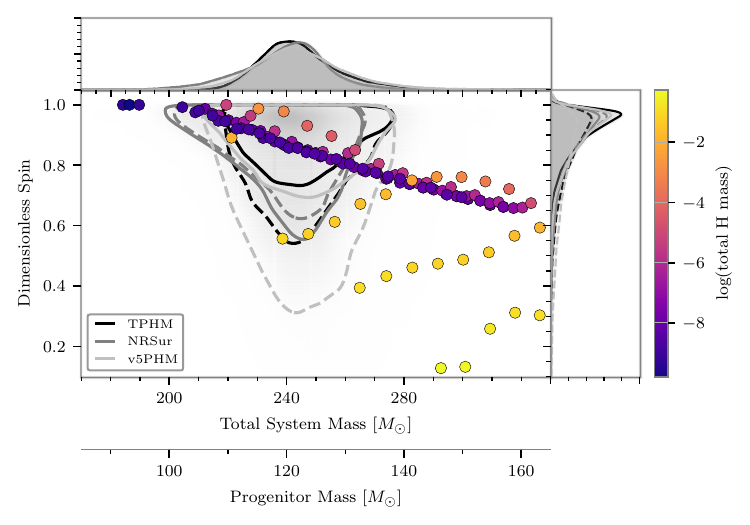}
    \caption{Dimensionless spin and final masses of our CHE progenitors, colorcoded by their final total hydrogen mass. The x-axis shows both the total mass of the binary system and the component masses assuming a mass ratio of $q=1$. The background contours indicate 90 \% credible intervals of the total BH masses and individual spins inferred for GW231123 from\cite{GW231123}, for three waveform models TPHM, NRSur, v5PHM. The solid (dashed) lines indicate the spin of the primary (secondary) BH. A clear downward trend is visible for our CHE models, the outliers being models that still have some hydrogen left in their envelopes.}
    \label{fig:kerr_final_mass}
\end{figure*}

\textit{Grid Setup.} 
We compute a grid of chemically homogeneously evolving (CHE) binary models with initial masses of $M_i = 120 - 180 \, M_\odot$ and initial orbital periods of $P_i = 0.7 - 1.8$ days at a metallicity of $Z = 10^{-5}$.  All models presented in the main figures are calculated assuming an initial mass ratio of unity, $q=M_1/M_2 = 1$, producing black hole binaries with equal masses, consistent with GW231123.  

We also explore test simulations where we vary the initial mass ratios and find qualitatively the same behavior. The tightest systems in our grid that are in overcontact at the start of the simulation tend to evolve towards equal masses \citep{Marchant_2016} and our models models assuming initial mass ratios of 1 describe these well. However most CHE systems are initially not in over contact. Starting with a mass ratio different from one, will also leads to black hole progenitors with systems different from one.  We leave a full exploration of this to future work. 

We interrupt the evolution of systems that either experience L2 overflow at onset of hydrogen burning or exhibit a difference between the central and surface helium abundance exceeding 0.2, indicating departure from CHE. Systems that remain chemically homogeneously are evolved until core helium depletion.

\section{Results}
\label{sec:results}

\subsection{Overview of the grid of progenitors}

In Fig. \ref{fig:grid_final_mass}, we show our computed grid, indicating the different evolutionary pathways of the systems. Systems undergoing chemically homogeneous evolution (CHE) are color-coded according to their final mass. 
At the very low metallicity we considered here, the CHE window is limited to a relatively limited range of orbital periods. At larger periods, rotational mixing becomes inefficient and the systems evolve out of CHE, while at shorter periods, the stars are so close that they overflow their outer Lagrangian point (L2) during the main sequence, potentially leading to a stellar merger. The window widens at higher masses, since stellar winds help remove the outer layers that are not well mixed. The tightest systems in the CHE window, marked with white dots, experience an overcontact phase at the onset of hydrogen burning. However, these constitute a minority of the systems in the CHE window. Most CHE systems are detached and stay detached. We further indicate with white crosses the systems with final helium core masses that are expected to lead to pair-instability supernova, leaving no remnant behind.  

For reference, we highlight with black squares systems whose final helium core masses and spins match the inferred total mass of GW231123' progenitors, $238^{+28}_{-49}\,M_\odot$, and are consistent with either the primary or secondary black hole spins within a 90\% credibility range. However, we remind the reader that the masses and spins of our models provide only an upper limit to the final black hole mass and angular momentum.

In Fig. \ref{fig:kerr_final_mass}, we compare the final total masses and the dimensionless spin of our progenitors with the inferred properties of GW231123’s progenitors \citep{GW231123}. We show the results from the RPhenomTPHM (TPHM), NRSur7dq4 (NRSur) and NRv5PHM (v5PHM) waveform models. The RPhenomXPHM and IMRPhenomXO4a are not shown here, as they exhibit notable differences relative to the other models (see Fig. 7 of \cite{GW231123}). Our results show that several of our models at the end of helium burning match the inferred properties of GW231123' progenitors, with a total system mass of $240 -266\, M_\odot$ and high individual spins in the range 0.8 - 0.9. 

Fig.~\ref{fig:kerr_final_mass} further shows that the large majority of our models show a tight downward correlation between dimensionless spin and final mass. This may be surprising as they originate from a range of initial masses and spins, and indicates the presence of a mechanism that converges the final spins to this relation. This trend is well described by $a_\mathrm{}\propto M^{-0.9} $ (see \hyperref[app:A]{Appendix A} for a derivation of the slope, c.f. \cite{Marchant_2024}) and is due to the stars reaching critical rotation as we will show in Sect.~\ref{sec:time_evolution}. The outliers from the trend are represented by systems that did not evolve fully chemically homogeneously and retained a small amount of hydrogen in their outer layers, as shown by the color bar. This affects their sizes and critical rotational velocities.

\begin{figure}[]
    \hspace*{-0.7cm} 
    \includegraphics[width=0.5\textwidth]{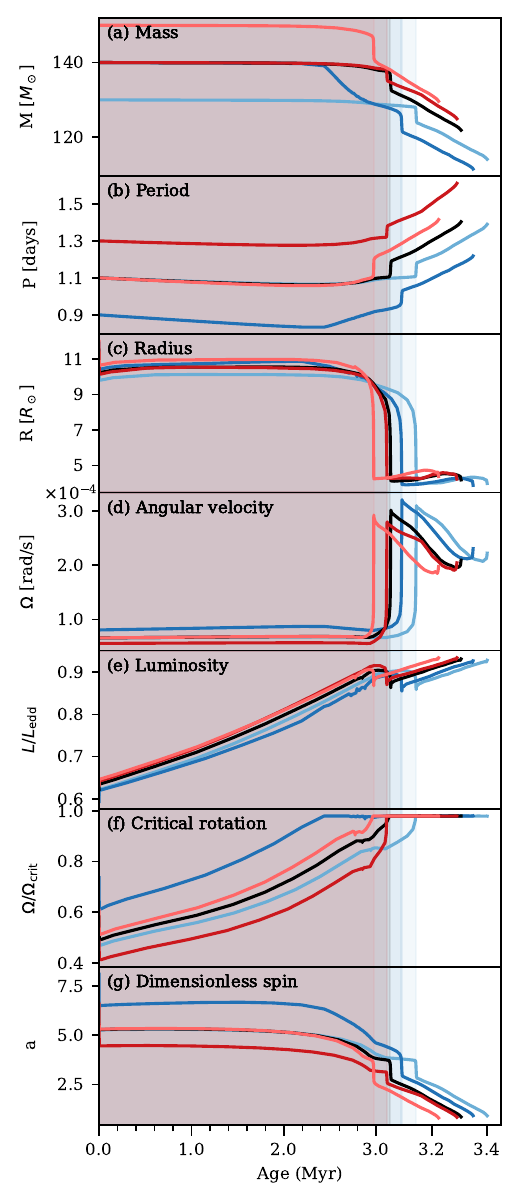}
    \caption{Evolution of CHE systems with model properties shown in Table~\ref{tab:candidates_tracks}. The main sequence is indicated by the shaded region. The x-axis is rescaled for better visualization of the late evolutionary stages.}
    \label{fig:time_evolution}
\end{figure}

\begin{deluxetable}{ccccc}
\tabletypesize{\scriptsize}
\tablewidth{0pt}
\tablecaption{Properties of the models shown in Fig.~\ref{fig:time_evolution}, with row colors matching the corresponding evolutionary tracks. \label{tab:candidates_tracks}}
\tablehead{
\colhead{ Initial mass } & \colhead{ Initial period} &  \colhead{Final mass} & \colhead{Kerr}     & \colhead{Merger time}  \\
[-1em] 
 \colhead{[$M_\odot$]}& \colhead{[days]} &  \colhead{[$M_\odot$]} & \colhead{-}  & \colhead{[Myr]} 
}
\startdata
            \rowcolor{lightBlue!50}
            130      &         1.1  &      114  &         0.917 &  53\\
             \rowcolor{darkBlue!60}
            140      &         0.9  &      111  &         0.920 &  39\\
            \rowcolor{midBlack!40} 
            140      &         1.1  &      122  &         0.859 &  49\\
             \rowcolor{darkRed!60}
            140      &         1.3  &      125   &        0.840 &  68\\
             \rowcolor{lightRed!50}
            150      &         1.1  &      130  &         0.805 &  45\\
\enddata
\end{deluxetable}

\subsection{Evolution of individual progenitor systems}
\label{sec:time_evolution}

In Fig.~\ref{fig:time_evolution}, we show the time evolution of key parameters for five example models with different initial masses and initial periods as shown in Table~\ref{tab:candidates_tracks}.  The properties of the examples models are chosen to be centered around a potential progenitor system of GW231123 with final total system mass of $244\, M_\odot$ and dimensionless spins of the individual stars of $ 0.86$. 

The first two panels of Fig.~\ref{fig:time_evolution} show the mass and the period evolution of the models. Due to the reduced winds at low metallicity, the stars do not loose significant mass during the main-sequence. However, once they reach central hydrogen depletion and contract into helium stars (panel c), they reach critical rotation (panel f and Eq.~\ref{eq:omega_crit}) and mass is shed, leading to the significant decrease in mass during the late evolutionary stages. The system that starts in the tightest configuration (dark blue) reaches critical rotation earliest, and as a result, experiences the most dramatic mass loss. Mass loss also leads to angular momentum loss from the system, which leads to widening of the orbit in our models, as can be seen in the increase in the orbital periods in panel b. 

Critical rotation is reached because of the spin-up during contraction after the main sequence ($\Omega \nearrow$, panel d), but mainly as a result of the steadily increasing luminosity throughout the evolution (panel e), which approaches the Eddington limit in the late stages. The increase in $L/L_\mathrm{edd}$, which corresponds to the $\Gamma$ factor in Eq.~\ref{eq:omega_crit}, lowers the critical rotation threshold ($\Omega_\mathrm{crit} \searrow$), driving the stars to reach the critical rotation limit ($\Omega/\Omega_\mathrm{crit} \nearrow 1$).

In the bottom panel g, we show the dimensionless spin (Eq.~\ref{eq:a_BH}). The initial angular momentum contained in the stellar progenitors leads to dimensionless spins well above 1 during the main sequence. As the stars evolve, the loss of angular momentum due to (rotationally-enhanced) winds reduce the spins drastically. Due to the limitation imposed by the stars' critical rotation, all the dimensionless spins are reduced to near or below 1 by the end of helium burning following a tight correlation between mass and spin as we showed already in Fig.~\ref{fig:kerr_final_mass}. This trend is the direct result of the critical rotation rate setting a maximum to the final progenitor's spin. 

Lastly, we compute the delay time until the black hole merger using the formula from \cite{Peters_1964} for the case of a circular orbit and show that all systems are expected to merge within a Hubble time, with typical prompt merger times for the systems calculated here (see Table \ref{tab:candidates_tracks}).  At higher metallicity and higher masses, not shown here, we expect the window for CHE evolution to widen and also lead to wider final progenitors with less prompt mergers.

\section{Discussion}
\label{sec:discussion}

\textit{Fate and Properties of the Remnants.}
The collapse of a very massive and rapidly rotating progenitor into a black hole is highly non trivial. Centrifugal support in the outer stellar layers, may prevent their immediate collapse. These regions could instead form an accretion disk around the black hole, which may not be fully accreted, thereby reducing the mass of the final remnant and altering its final spin \citep[e.g.][]{Beloborodov_1999, Yoon_Langer_2005, Fuller_2022}. This is certainly a viable way to create black holes from rapidly rotating progenitors with final masses in the hypothesized pair instability mass gap \citep{Sieger_2022}. 


In the context of GW231123, \cite{Gottlieb_2025} recently modeled the collapse of a low-metallicity, rotating, magnetized helium core above the mass gap using MESA and general-relativistic magnetohydrodynamic codes. Their helium star model has a mass of $140\, M_\odot$ at collapse, comparable with some of our progenitor models. They show that the magnetic field strength plays a crucial role in the final outcome. In the case of weak magnetic fields, no outflows are present, so the black hole accretes almost the full stellar mass, $M_\mathrm{BH} \approx M_\mathrm{He}$, and reaches maximum spin, $a \approx 1.0$, consistent with the primary BH of GW231123. For intermediate magnetic fields, outflows are initiated only after most of the mass has been accreted, producing a $100\, M_\odot$ BH within the mass gap with a spin of $a\approx 0.8$, matching the properties of the secondary BH. Their results suggest that at least for weak magnetic fields, the assumption of direct collapse with full conservation of mass and angular momentum that we are effectively using might be reasonable. Moreover, our CHE systems are promising candidates for the origin of the high-mass, rapidly spinning progenitor described by \cite{Gottlieb_2025}, and may naturally explain the formation of GW231123.

However, \cite{Croon_2025} raises the question whether such highly spinning progenitor stars would form black holes at all. While non-rotating models above $>120\, M_\odot$ are expected to undergo direct collapse due to photodisintegration, rotation might shift this upper edge to higher values. The central temperature of rapidly rotating stars is lower than in non-rotating models due to the reduction of the effective gravity. Consequently, they may not reach the conditions required for the onset of the photodisintegration and direct collapse. Instead, pair production may trigger a complete disruption of the star, leaving no remnant behind. \cite{Croon_2025} shows that, in their models, a pair-instability supernova takes place for a $160\, M_\odot$ helium star rotating faster than $>0.4\cdot \Omega_\mathrm{crit}$ ($>0.5\cdot \Omega_\mathrm{crit}$) for a median ($+3\sigma$) $^{12}\mathrm{C}(\alpha,\gamma)^{16}\mathrm{O}$ rate. This confronts us with the possibility that our progenitors, as well as the progenitor in \cite{Gottlieb_2025}, might not collapse into a black hole. This may be due in part to the fact that \cite{Croon_2025}, in contrast to our work and \cite{Gottlieb_2025}, does not account for angular momentum transport through magnetic fields, which leaves the interiors more rapidly spinning, see also \citet{Marchant_2020}.  Further investigation is required to asses the fate of such rapidly rotating, massive stars above the upper mass gap.

\textit{Spin-orbit Misalignment.}
Our models can reproduce the high individual spins of GW231123's progenitors. However, the relatively low effective inspiral spin $\chi_\mathrm{eff} = 0.31^{+0.24}_{-0.39}$ \citep{GW231123}, indicates some degree of spin-orbit misalignment, which contrasts with the naive expectation for direct complete collapse, which would produce fully aligned spins. Nonetheless, slight misalignment can be conceivable due to asymmetric fall-back during collapse. We also note that the inferred $\chi_\mathrm{eff}$ value is highly sensitive to the spin orientations, which are not very well constrained. LIGO Hanford shows support for aligned spins, while LIGO Livingston favors misalignment \citep{GW231123}. 

\textit{The Role of Tides.}
In our models, tides efficiently synchronize the stellar spins with the orbit during the main sequence, but are assumed to become negligible once the stars contract into helium stars. This assumption is motivated by the strong dependence of the tidal timescale on the stellar radius. Without the effect of synchronization after the main sequence, the stars rapidly spin during the contraction, leading to almost maximally spinning black holes. In contrast, the CHE models above the pair instability mass gap presented by \cite{Marchant_2016} and \citet{Buisson_2020} exhibit much lower Kerr parameters. This is  because, in their models, tidal synchronization is maintained after the main-sequence, causing the stars to spin down. As already noted by \citet{Marchant_2024} this may be spurious in their simulations.

\textit{Expected Rates.}
Our grid focuses on very massive stars with initial masses in the range $M = 120 - 180 \, M_\odot$. The existence of such massive stars is supported by observations \citep[e.g.\ ][]{Crowther_2016}. Observations also indicate that the close binary fraction increases with stellar mass \citep{Sana_2012, Offner_2023} and this trend extends to low-metallicity environments \citep{Sana_2025}. However, our understanding of the initial mass function (IMF) and very close binary fraction at low metallicity remains highly uncertain. Estimating rates for such extreme masses and short-period systems would therefore require significant extrapolation, making any quantitative predictions of limited reliability. Moreover, the progenitors we consider here may form in dense stellar systems where the most massive stars sink to the center where they can potentially pair. For this reason, we refrain from providing any rate estimates in this work.


\section{Conclusions}
\label{sec:conclusions}
We have investigated the chemically homogeneous evolution (CHE) channel for the formation of massive, rapidly rotating black-hole progenitors with component masses above the pair-instability gap. 
Within our grid of models evolved until the end of helium burning, several models reproduce the properties of GW231123-like progenitors, forming systems with a total mass of $240 - 266\, M_\odot$ and high individual spins of $a=0.8-0.9$. 
This supports CHE as a viable formation channel for this event, although the final collapse and the resulting spins are considerably uncertain.

We further find a remarkably tight correlation between the dimensionless spins and final masses, $a\propto M^{-0.9}$, for models that have no hydrogen left in their envelopes. We show that this trend can be understood as a direct consequence of the critical rotation rate (Eq.~\ref{eq:omega_crit}), which sets a maximum at which the progenitor star can rotate, see also \citet{Marchant_2024}. This limit is easily reached in our progenitors when they spin up contracting after completing the main sequence, since their luminosities are close to their Eddington luminosities.

\begin{acknowledgments}
The authors are particularly grateful to Jim Fuller for extensive discussions of this project.  We also acknowledge Jakob Steggmann, Alejandro Vigna-Gomez, Aleksandra Olejak, Stephen Justham, Ruggero Valli, Jing-Ze Ma, Diego Calderon, Lieke van Son and Andrei Belobodorov for their help and insightful discussions at various times.   
\end{acknowledgments}

\section*{Appendix}
\label{sec:Appendix}

\section*{A: Correllation between the Progenitor's  Dimensionless Spins and Masses }
\phantomsection
\label{app:A}

In Fig.~\ref{fig:kerr_final_mass} we find a strong correlation between the final dimensionless spin and the mass for progenitors in our grid that have no hydrogen left in their envelopes. Inspection of the  evolution of individual progenitors, in Fig.~\ref{fig:time_evolution}, suggested that the origin for the tight final relation is the limiting effect of the critical rotation rate (Eq.~\ref{eq:omega_crit}) for the progenitor stars.  Here we show how this limiting effect leads to a tight and well-defined relation in the final spin with mass.

\begin{figure}
    \centering
    \includegraphics[]{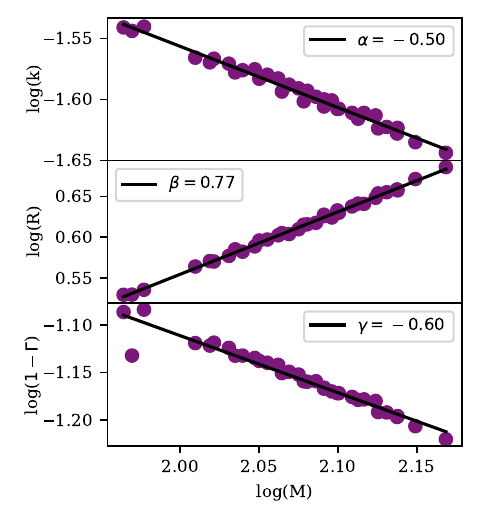}
    \caption{Scaling relations for our critically rotating helium star models, showing the dependence of the gyration constant (top), stellar radius (middle) and one minus the Eddington factor (bottom) on mass. The linear fits are used to determined the exponents $\alpha$, $\beta$ and $\gamma$ from Eq.~\ref{eq:alpha_beta}.}

    \label{fig:scaling}
\end{figure}

\begin{figure}
    \centering
    \includegraphics[]{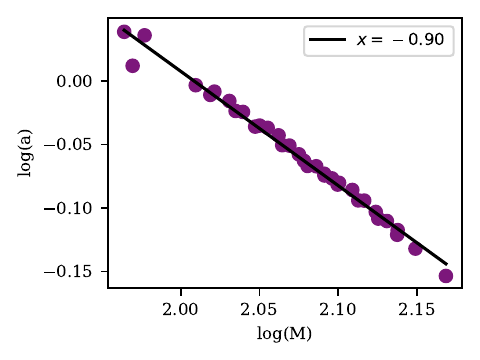}
    \caption{ Same as Fig.~\ref{fig:kerr_final_mass} in the main text, but using logarithmic axis and only showing the fully chemically homogeneous models which have no more H left in their envelopes. We find that these models follow a very tight  $a \propto M^{-0.9}$, which can be understood to be the result of the limiting effect of the critical rotation of the progenitor.
    \label{fig:powelaw_a_mass}}
\end{figure}

The angular momentum of a critically rotating star is given by:
\begin{equation}
    J_\mathrm{crit} = I\Omega_\mathrm{crit} = kMR^2\sqrt{\frac{GM(1-\Gamma)}{R_\mathrm{eq}^3}}\, ,
\end{equation}
where $I$ is the moment of inertia, $k = (\beta/R)^2$ the gyration constant defined as the squared ratio of the gyration radius $\beta$ to the stellar radius $R$, $\Gamma = L/L_\mathrm{edd} \propto L/M$ the Eddington factor, $M$ the stellar mass and $R_\mathrm{eq} = 1.5\, R$ the equatorial radius. Substituting this in Eq.~\ref{eq:a_BH}, the dimensionless spin parameter of a critically spinning star becomes:

\begin{equation}
    a_\mathrm{crit} = \frac{c\,J_\mathrm{crit}}{G\,M^2} \propto k \, \sqrt\frac{R(1-\Gamma)}{M}\,.
\end{equation}
Assuming the scaling relations $k\propto M^\alpha$, $R\propto M^\beta$ and $1-\Gamma  \propto M^\gamma$, this expression reduces to:

\begin{equation}
     a_\mathrm{crit} \propto M^{(\beta+\gamma-1)/2 +\alpha}\,.
     \label{eq:alpha_beta}
\end{equation}

For our helium stars, we find $\alpha=-0.5$ $\beta = 0.77$ and $\gamma = -0.60$ (see Fig.~\ref{fig:scaling}), which  leads to:
\begin{equation}
    a_\mathrm{crit} \propto M^{-0.9}\, ,
\end{equation}
which corresponds to the downward correlation between mass and spins shown in Fig.~\ref{fig:kerr_final_mass} and Fig.~\ref{fig:powelaw_a_mass}.

We note that this explanation and the derivation is largely identical to the one provided in \citet{Marchant_2024}
when they discuss a semi-analytical model for the spins of black hole binaries.  We find the same downward trend with mass.  However, the quantitative predictions they present based on this differ significantly from our findings.   In their work they predict very massive stars above the PISN gap to be slowly spinning, producing black holes with Kerr parameters $a \lesssim 0.25$, see their Fig.~5, while we find dimensionless spins reaching up to 1.   

We expect that the reason for the difference is that \citet{Marchant_2024} use power-law relations derived from single star models that have been artificially forced to evolve chemically homogeneously through all evolutionary stages up to carbon depletion. This is a rather extreme assumption in their simulations. In our simulations we find that the progenitors are only homogeneous during central H burning, but develop internal composition gradient during core helium burning. Even though we do not compute further, the mixing timescales are not short enough to keep the star homogeneous in the later burning stages, which is not accounted for in their semi-analytical model. Probably as a results of this, they have very compact sizes for their final progenitors as can be inferred from their Fig.~1. 
In their semi-analytical model they further assume the gyration constant $k$ to be independent of mass and evolutionary state, which may not be very realistic.

\begin{figure}
    \centering
    \includegraphics[]{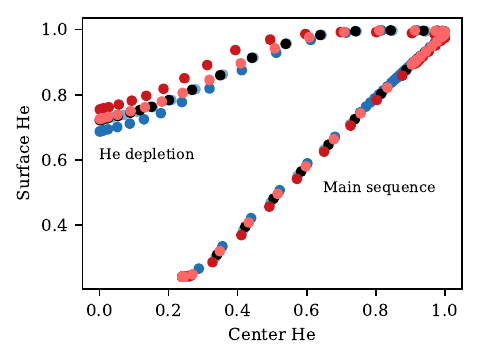}
    \caption{Surface vs center helium evolution for the models shown in Fig.~\ref{fig:time_evolution} using the same colors. Even though the models are chemically homogeneous on the main-sequence, the surface He abundance starts to diverge from the core He abundance during helium burning.   }
    \label{fig:surface_core_He}
\end{figure}

\section*{B: Additional Plots for the example models}
In Fig.~\ref{fig:surface_core_He} we show the evolution of the surface helium abundance versus the central helium abundance for the example models discussed in Sect.~\ref{sec:results}. The models stars with a central and surface helium abundance of 0.24 at the bottom of the diagram.  During the main sequence the stars stay fully homogeneous with central and surface helium abundance both steadily increasing to 1.0.  After the main sequence, when helium ignites in the center, the stars no longer evolve chemically homogeneously. The central helium abundance drops, while the surface helium abundance stays above 0.7.

\section*{C: Progenitor properties}
In Table~\ref{tab:candidates} we show the properties of all the models from our grid that match the final black hole mass and individual spins of GW231123's progenitors.


%

\begin{deluxetable}{ccccc}[h!]
\tabletypesize{\scriptsize}
\tablewidth{0pt}
\tablecaption{Properties of models consistent with GW231123's progenitors. \label{tab:candidates}}
\tablehead{
\colhead{ Initial mass } & \colhead{ Initial period} &  \colhead{Final mass} & \colhead{Kerr}     & \colhead{Merger time}  \\
[-1em] 
 \colhead{[$M_\odot$]}& \colhead{[days]} &  \colhead{[$M_\odot$]} & \colhead{-}  & \colhead{[Myr]} 
}
\startdata
     135       &     1.3    &    121      &    0.88     &      71\\
     135       &     1.4    &    122      &    0.86     &      82\\
     135       &     1.5    &    124      &    0.93     &      91\\
     140       &     1.1    &    122      &    0.86     &      49\\
     140       &     1.2    &    124      &    0.85     &      58\\
     140       &     1.3    &    125      &    0.84     &      68\\
     140       &     1.4    &    126      &    0.84     &      78\\
     140       &     1.5    &    128      &    0.90      &      88\\
     145       &     1.0      &    123      &    0.85     &      39\\
     145       &     1.1    &    126      &    0.83     &      47\\
     145       &     1.2    &    128      &    0.82     &      56\\
     145       &     1.3    &    129      &    0.81     &      65\\
     150       &     1.0      &    126      &    0.83     &      39\\
     150       &     1.1    &    130      &    0.81     &      45\\
     155       &     1.0      &    128      &    0.82     &      39\\
     165       &     0.9    &    120      &    0.86     &      39\\
     170       &     0.9    &    122      &    0.86     &      39\\
     175       &     0.9    &    123      &    0.84     &      39\\
     180       &     0.9    &    125      &    0.84     &      39\\
     135       &     1.6    &    124      &    0.57     &     108\\
     140       &     1.6    &    128      &    0.61     &     102\\
\enddata
\end{deluxetable}

\bibliography{ref}
\bibliographystyle{aasjournalv7}



\end{document}